Stress hormones predict hyperbolic time-discount rates six months later in adults.

Taiki Takahashi[1], Mizuho Shinada[1], Keigo Inukai[1,2], Shigehito Tanida[1], Chisato Takahashi[1], Nobuhiro Mifune[1,2], Haruto Takagishi[1,2], Yutaka Horita[1,2], Hirofumi Hashimoto[1,2], Kunihiro Yokota[1], Tatsuya Kameda[1], Toshio Yamagishi[1]

[1] Department of Behavioral Science, Hokkaido University

[2] Japan Society for the Promotion of Sciences

Corresponding Author: Taiki Takahashi

Email: taikitakahashi@gmail.com

Department of Behavioral Science,  Hokkaido University

N.10, W.7, Kita-ku, Sapporo, 060-0810, Japan

TEL: +81-11-706-3057     FAX: +81-11-706-3066

Summary

OBJECTIVES: Stress hormones have been associated with temporal discounting. Although time-discount rate is shown to be stable over a long term, no study to date examines whether individual differences in stress hormones could predict individuals' time-discount rates in the relatively distant future (e.g., six month later), which is of interest in neuroeconomics of stress-addiction association. METHODS: We assessed 87 participants' salivary stress hormone (cortisol, cortisone, and alpha-amylase) levels and hyperbolic discounting of delayed rewards consisting of three magnitudes, at the time-interval of six months. For salivary steroid assays, we employed a liquid chromatography/ mass spectroscopy (LC/MS) method. The correlations between the stress hormone levels and time-discount rates were examined. RESULTS: We observed that salivary alpha-amylase (sAA) levels were negatively associated with time-discount rates in never-smokers. Notably, salivary levels of stress steroids (i.e., cortisol and cortisone) negatively and positively related to time-discount rates in men and women, respectively, in never-smokers. Ever-smokers' discount rates were not predicted from these stress hormone levels. CONCLUSIONS: Individual differences in stress hormone levels predict impulsivity in temporal discounting in the future. There are sex differences in the effect of stress steroids on temporal discounting; while there was no sex defference in the relationship between sAA and temporal discounting.

# 1. Introduction

Because impulsivity is a core deficit in neuropsychiatric disruptions such as addiction, ADHD, depression, and psychopathy, neurobiological bases of impulsivity operationalilzed as "delay discounting" (preference for smaller sooner rewards over larger later ones) have extensively been investigated in neuroeconomics (see Takahashi, 2009, for a review). In order to test whether impulsivity operationalized as above is stable over a long time-interval or not, Ohmura et al (2006) in our group and Kirby (2009) examined the stability of the time-discount rate (in addition to other discounting parameters). The authors reported that temporal discounting is a stable trait over several months to one year. This suggests impulsivity in intertemporal choice (the degree to which delayed rewards are discounted) might be determined by considerably stable neurobiological machineries.

Regarding neurobiological substrates which modulate temporal discounting, Takahashi's group has extensively investigated the roles of hormones (stress/sex steroids and amylase) in intertemporal choice (Takahashi, 2005; Takahashi et al., 2006; Takahashi et al., 2007). The previous studies by Takahashi and colleagues demonstrated that stress hormones (i.e., cortisol and salivary alpha-amylase) are negatively associated with impulsivity in intertemporal choice in young never-smoking men (university students) (Takahashi, 2004; Takahashi et al., 2007). Regarding stress systems, two primary systems are particularly involved in setting on the stress response, hypothalamus—pituitary—adrenocortical axis (HPA) and sympatho-adrenomedullary (SAM) system. The salivary enzyme alpha-amylase has been proposed as a marker for stress-induced activity of the sympathetic nervous system (SNS) and recent studies have underscored the usefulness of salivary alpha-amylase (sAA) in this regard (Nater et al., 2009). It is widely accepted that psychological stress could produce physiological effects. Two primary systems are particularly involved in setting on the stress response, hypothalamus—pituitary—adrenocortical axis (HPA) and sympatho-adrenomedullary (SAM) system. The activation of HPA causes an increase in cortisol secretion in the adrenal cortex; while activation of SAM causes an increase in sAA (Nater et al., 2009). Because temporal discounting and its relations to hormones are influenced by smoking status (Ohmura et al., 2005), sex (Lucas et al., 2010), and age (Read et al., 2004), it is important to examine whether the previous findings in young never-smoking men can be generalized into broader populations. Also, because temporal discounting is a stable trait, it is important to examine whether individual differences in the levels of stress hormones (i.e., cortisol, cortisone, and sAA) can predict subjects' impulsivity in the distant future (e.g., six months later), in order to establish the medical methods of diagnosis and prediction of future neuropsychiatric disruptions associated with impulsivity (e.g., addiction). It is also to be noted that this is the first study to examine the role of cortisone in intertemporal choice, which is of importance in neuroeconomics and endocrinological economics.

## 2. Methods

### 2.1. Subjects

Participants (N=87) were recruited through advertisements run in local newspapers. The mean age of participants was 47.3 years (range = 21-69). The group of participants consisted of 49 never-smokers (14 male never-smokers and 35 female never-smokers) and 38 ever-smokers (28 male ever-smokers and 10 female ever-smokers). The ever-smokers consisted of current and ex- smokers, but there was no significant differences regarding the effects of stress hormones on temporal discounting between current and ex- smokers, we combined both current and ex- smokers into the ever-smokers. Because sex and smoking status are associated with neuroendocrinological substrates (e.g., cortisol) which mediate temporal discounting (Mello, 2009; Takahashi, 2004), we mainly examined the relationships between stress hormones and time-discount rates within the divided four groups (i.e., male never-smokers, female never-smokers, male ever-smokers, and female ever-smokers).

### 2.2. Experimental protocols

We collected all participants' saliva samples between 9:00 and 9:30AM for the assessment of stress hormones (i.e., salivary cortisol, cortisone, amylase). Participants were given instructions not to eat or drink anything except for water, and to refrain from physical exercise before the experiment. For the assessment of salivary alpha-amylase level, we collected three salivary alpha-amylase samples from each participant, and calculated the mean of all three samples.

### 2.3. Assessments of salivary stress hormones

In order to measure salivary stress steroids (i.e., cortisol and cortisone), we employed the same saliva collecting procedures as in our previous studies (Takahashi et al., 2006). After collecting participants' saliva, the saliva samples were immediately frozen and kept at -20$^{o}$C until hormonal assays. Salivary cortisol and cortisone levels were assessed with a liquid chromatography/mass spectroscopy (LC/MS) method. All procedures for steroid hormonal assays (i.e., cortisol and cortisone) were conducted at ASKA pharmaceutical company (Tokyo, Japan) which has significant experience in hormonal assays (Takahashi, 2004; Takahashi, et al. 2006). Staff at the company did not know the nature of our present study.

To measure SAM system activity, we measured sAA with the same methodology and device (amylase-monitor, Nipro Co. Ltd, Japan) as in our previous studies (Takahashi et al., 2007; Takagishi et al. 2009). This device, developed by analytical chemists Yamaguchi and colleagues (2006), utilizes a reagent paper containing 2-chloro-4-O-beta-d-galactopyranosylmaltoside (Gal-G2-CNP), a substrate of amylase. When Gal-G2-CNP is hydrolyzed by amylase, the hydrolyzed product (CNP) changes emission wavelengths (reflectance) with time. The collecting paper was directly inserted into the oral cavity, and approximately 20-30microL of saliva was collected from under the tongue over a period of 10 to 30 seconds. The reflectance 30s after the initial time was automatically measured by the optical device. In total, the measurement of sAA level was completed in approximately one minute. The levels from the three sAA levels of the three samples were significantly correlated with each other ($ps < .0001$).

**2.4. Kirby MCQ**

Six months after the collection of saliva samples, the subjects' hyperbolic discounting rates were assessed. We adopted the same procedure for assessing subject's discount rates as previous behavioral and neuroendocrinological studies of intertemporal choice (Kirby et al., 1999; Takahashi et al., 2006; Takahashi et al., 2007; Kirby, 2009). Studies in neuropsychopharmacology, psychoneuroendocrinology, and behavioral neuroeconomics have repeatedly observed that human and animal subject's delay discounting is well described by the hyperbolic discount function (see Takahashi, 2009, for a review):

$$V(D) = 1/(1+kD) \text{ (equation 1)}$$

where V(D) is a subjective value of delayed rewards at delay time D, and k (a hyperbolic discount rate) is a free parameter indicating subject's impulsivity in intertemporal choice (larger k values correspond to more rapid discounting; while smaller k values indicate self-control in intertemporal choice). In order to assess subject's discount rate k, as defined in equation 1, Kirby's MCQ (Kirby et al., 1999) was used. Kirby's MCQ consists of 27 questions relating to a choice between smaller immediate rewards and larger but delayed rewards (e.g. "Would you prefer 54 dollars today or 55 dollars in 117 days?"). According to the standard analysis procedure of MCQ, established by Kirby and colleagues (Kirby et al., 1999), we calculated subjects' discounting rates (i.e. ks) of three different sizes (small, medium, and large) of monetary gains. A total of three discount rates (i.e., small, medium, and large gains) were obtained for each subject (=k(S), k(M), and k(L), respectively). Geometric-mean discounting rates for different sizes (=k (mean)) were calculated, following Kirby's procedure (Kirby et al., 1999; Kirby, 2009). We then examined relationships between the hyperbolic time-discount rates of gains and stress hormone levels. In our Kirby MCQ form, all gains were expressed in terms of Japanese yen, with an exchange rate of one dollar to 100 yen. Because the distribution of the discount rate k is known to be skewed, we used logged k in the following analysis, according to a standard analytical procedure (Kirby et al., 1999; Kirby, 2009).

## 3. Results

### 3.1. Salivary stress hormones in male and female ever-smokers and never-smokers

Table 1 shows salivary levels of stress hormones (alpha-amylase, cortisol, cortisone) for the divided, upon sex and smoking status, four groups: male never-smokers, female never-smokers, male ever-smokers, and female ever-smokers. Data are expressed as the Mean (SD) in Table 1. There was no significant difference in stress hormone levels with respect to sex and smoking status. Salivary cortisol and cortisone levels were positively correlated in box sexes (male: $r=0.93$; female: $r=0.75$, all $ps<.0001$); while sAA did not relate to corticosteroids ($p>0.1$).

<insert Table 1 here>

There was a significant correlation between age and sAA ($r= 0.33$, $p<.01$ ); while there was no significant correlation between cortisol/cortisone levels and age ($ps>0.3$). After controlling for age, essentially the same results as below were obtained for the relationships between stress hormone levels and discount rates. It is also to be noted that there was no relationship between age and discount rates in our present sample (all ps>.5).

**3.2. Salivary stress hormones and time-discount rates in male and female never- and ever-smokers**

We first examined the relationships between stress hormone levels and discount rates in never-smokers, because smoking may influence hormones which have potential effects on temporal discounting behavior (Mello, 2009; Takahashi, 2004). Among stress hormones, we observed that sAA level was significantly and negatively correlated with logged time-discount rates in the never-smoker group, consistent with our previous study employing male never-smokers and demonstrated negative correlations between sAA and time-discount rates assessed within the same day (Takahashi et al., 2007). The Pearson's correlation coefficients for the never-smokers are presented in Table 2. The corresponding scatterplot between sAA and time-discount rates is presented in Figure 1. These correlations indicate that sAA levels strongly associate with never-smokers' discount-rates which were assessed six month later, in a negative manner. It is also to be noted that there was no sex difference in these negative relationships, as discussed later.

<insert Table 2 here>

<insert Fig.1 here>

Next, we divided the never-smokers into subgroups (i.e., male and female never-smokers). We then conducted Pearson's correlation analyses between stress hormone levels and logged time-discount rates in the two subgroups separately. For male never-smokers, there were significant negative correlations between stress hormones and several types of time-discount rates (Table 3). The negative direction of the relationships between stress hormones and temporal discounting in male never-smokers are consistent with that in our previous studies (Takahashi, 2004; Takahashi et al., 2007). Interestingly, cortisone was a stronger predictor of male never-smokers' temporal discounting, in comparison to cortisol. On the other hand, for female never-smokers, sAA levels negatively related to time-discount rates; while stress corticosteroids (i.e., cortisol and cortisone) positively related to several types of time-discount rates (Table 4). This finding indicates that there are sex differences in corticosteroids' effect on temporal discounting; while salivary alpha-amylase do not differently predicts male and female temporal discounting behavior six months later.

<insert Table 3 here>

<insert Table 4 here>

Finally, we examined the relationships between stress hormones and temporal discounting in ever-smokers. We observed no relationship between stress hormones between temporal discounting in both male (Table 5) and female (Table 6) ever-smokers, implying that ever-smokers' temporal discounting behavior is not affected by or predicted from these stress hormones. Finally, it is to be noted that there was no significant difference in discount rates between ever-smokers and never-smokers in the present participants, consistent with our previous finding observed in mild smokers in Japan (Ohmura et al., 2005, according to the criterion in this previous study, only two smokers were heavy smokers in our present study)

## 4. Discussion

This study is the first to (i) demonstrate that salivary stress hormone levels (i.e., alpha-amylase, cortisol, and cortisone) predict hyperbolic time-discount rates assessed six month later in never-smokers, (ii) observe sex differences in the relationships between stress steroid hormones and hyperbolic discounting, (iii) cortisone is a strong predictor of male never-smokers' temporal discounting behavior in the future, and (iv) show that ever-smokers' hyperbolic discount rates cannot be predicted from their stress hormone levels. The finding that sAA negatively affected temporal discounting in both male and female never-smokers indicates that there is no sex difference in the role of SAM system in modulating impulsivity in intertemporal choice. With respect to the relationships between corticosteroids and temporal discounting, there were sex differences. The observed negative effects of stress hormones on temporal discounting by the male never-smokers are consistent with our previous studies (Takahashi, 2004; Takahashi et al., 2007). Although cortisone strongly associated with male never-smokers' temporal discounting in a negative manner,(Table 3) the negative relationships between cortisol and temporal discounting were weak (Table 3) in the present study. The reason for the weak effect of cortisol on temporal discounting in male never-smokers may be that we assessed participants' time-discount rates 6 months later than the assessment of salivary stress hormones, while our previous study (Takahashi, 2004) examined the effect of cortisol on temporal discounting within the same day. Also, in our previous study (Takahashi, 2004), saliva sampling for a cortisol assay was conducted in the afternoon when the influence of the diurnal rhythm on cortisol level is minimal; while, in the present study, saliva sampling for cortisol assays was conducted between 9:00-9:30AM (see the method section), which might have blunted the effects of individual differences in cortisol levels on temporal discounting. The findings that (i) the timing of saliva collecting in the present study reduces the effect of the circadian rhythm on sAA (Nater et al., 2007), and (ii) there were significant correlations between sAA and most temporal discounting parameters of male never-smokers (Table 3) support this interpretation. Interestingly, corticosteroids consistently affected several types of parameters of female never-smokers' temporal discounting in a positive manner (Table 4). This indicates that women with higher stress steroids (e.g. corisol and cortisone) are likely to make more impulsive decisions in intertemporal choice. Furthermore, the roles of HPA and SAM systems in temporal discounting processes may be opposite in women, but similar in men. The observed sex differences may reflect distinct patterns of glucocorticoid receptor-expression or dopamine metabolism in the dopaminergic systems such as the striatum . Taken together, future neuroeconomic studies should examine the roles of neurobiological sex differences in the effects of the stress hormones on temporal discounting behavior, for a better understanding of neurobiological mechanisms underlying temporal discounting processes.

Table 1 Stress hormone levels for male and female never- and ever- smokers.

| | Amylase [kU/L] | Cortisol [ng/mL] | Cortisone [ng/mL] |
|---|---|---|---|
| male never-smokers | 64.452 (63.303) | 1.903 (0.728) | 13.582 (3.108) |
| female never-smokers | 52.329 (38.835) | 1.644 (0.865) | 13.829 (4.974) |
| male ever-smokers | 52.917 (42.955) | 2.112 (1.568) | 15.494 (7.613) |
| female ever-smokers | 38.150 (32.663) | 1.947 (1.565) | 15.165 (6.462) |

There was no significant effect of sex and smoking status on stress hormone levels.

Table 2. Correlations between stress hormones and time-discount rates in a group consisting of male and female never-smokers.

|  | Amylase | Cortisol | Cortisone |
|---|---|---|---|
| Log k(mean) | -0.43* | 0.12 | 0.17 |
|  | (0.00) | (0.41) | (0.24) |
| Log k (S) | -0.45* | 0.05 | 0.05 |
|  | (0.00) | (0.75) | (0.72) |
| Log k(M) | -0.41* | 0.14 | 0.22 |
|  | (0.00) | (0.34) | (0.13) |
| Log k(L) | -0.41* | 0.17 | 0.20 |
|  | (0.00) | (0.25) | (0.17) |

There were negative correlations between sAA and temporal discounting in never-smokers. P-values are presented in the parentheses (*: $p < .05$).

Table 3. Correlations between stress hormones and time-discount rates in male never-smokers

|  | Amylase | Cortisol | Cortisone |
|---|---|---|---|
| Log k (mean) | -0.58* | -0.42 | -0.56* |
|  | (0.03) | (0.13) | (0.04) |
| Log k (S) | -0.65* | -0.38 | -0.52 |
|  | (0.01) | (0.18) | (0.06) |
| Log k (M) | -0.55* | -0.43 | -0.54* |
|  | (0.04) | (0.13) | (0.04) |
| Log k (L) | -0.40 | -0.46 | -0.58* |
|  | (0.18) | (0.12) | (0.04) |

All stress hormones were related to several types of time-discount rates in a negative manner in never-smoking men. P-values are presented in the parentheses (*: $p < .05$)

Table 4 Correlation between stress hormones and time-discount rates in female never-smokers

|  | Amylase | Cortisol | Cortisone |
|---|---|---|---|
| Log k(mean) | -0.38* | 0.29 | 0.37* |
|  | (0.03) | (0.09) | (0.03) |
| Log k (S) | -0.34 | 0.19 | 0.21 |
|  | (0.05) | (0.28) | (0.24) |
| Log k(M) | -0.37* | 0.30 | 0.42* |
|  | (0.03) | (0.08) | (0.01) |
| Log k(L) | -0.43* | 0.38* | 0.45* |
|  | (0.01) | (0.03) | (0.01) |

Salivary alpha-amylase levels negatively related to time-discount rates; while stress steroids positively related to several types of time-discount rates in never-smoking women. P-values are presented in the parentheses (*: $p < .05$)

Table 5 Correlation between stress hormones and time-discount rates in male ever-smokers

|  | Amylase | Cortisol | Cortisone |
|---|---|---|---|
| Log k(mean) | 0.23 | -0.19 | -0.15 |
|  | (0.25) | (0.34) | (0.44) |
| Log k (S) | 0.21 | -0.13 | -0.06 |
|  | (0.32) | (0.54) | (0.79) |
| Log k(M) | 0.26 | -0.25 | -0.21 |
|  | (0.19) | (0.21) | (0.29) |
| Log k(L) | 0.24 | -0.20 | -0.16 |
|  | (0.25) | (0.33) | (0.44) |

No significant relation between stress hormones and time-discount rates was observed. P-values are presented in the parentheses (*: $p < .05$)

Table 6 Correlation between stress hormones and time-discount rates in female ever-smokers

|  | Amylase | Cortisol | Cortisone |
|---|---|---|---|
| Log k(mean) | 0.40 | -0.41 | -0.55 |
|  | (0.25) | (0.25) | (0.10) |
| Log k (S) | 0.30 | -0.33 | -0.52 |
|  | (0.41) | (0.35) | (0.12) |
| Log k(M) | 0.54 | -0.41 | -0.49 |
|  | (0.11) | (0.24) | (0.15) |
| Log k(L) | 0.42 | -0.43 | -0.59 |
|  | (0.30) | (0.29) | (0.12) |

No significant relation between stress hormones and time-discount rates was observed. P-values are presented in the parentheses.(*: $p < .05$)

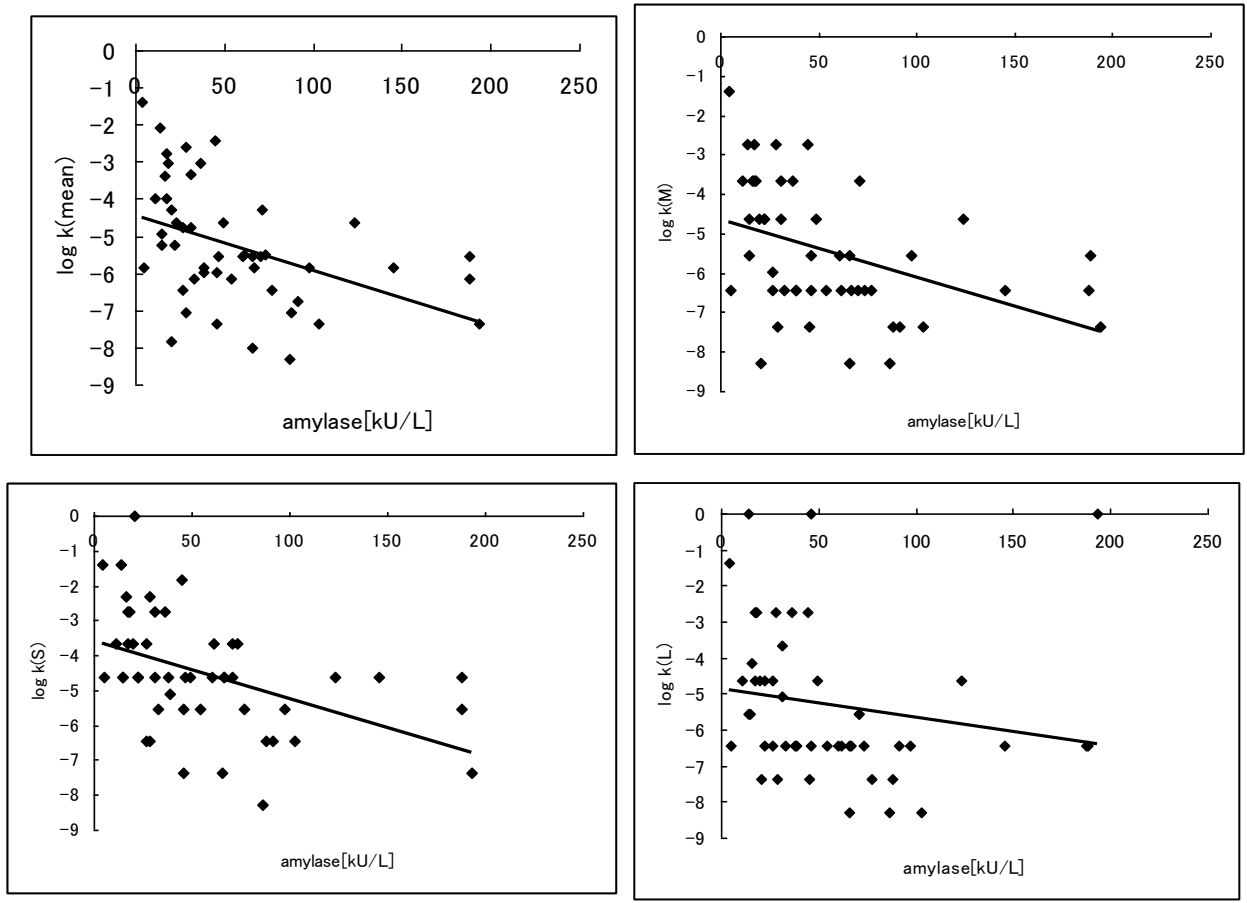

Figure 1. Scatterplot of sAA level and time-discount rates in never-smokers consisting of men and women. Significant negative relationships were observed. Note that there was no sex difference in these negative relationships (see the result section).